\documentclass{aa}
\usepackage{psfig,txfonts}

\begin{document}

\title{Serpens X-1 observed by {\it INTEGRAL}\thanks{Based on
observations with {\it INTEGRAL}, an ESA project
with instruments and science data centre funded by ESA member
states (especially the PI countries: Denmark, France, Germany,
Italy, Switzerland, Spain), Czech Republic and Poland, and with
the participation of Russia and the USA.}}

\author{N. Masetti\inst{1},
L. Foschini\inst{1},
E. Palazzi\inst{1},
V. Beckmann\inst{2,3},
N. Lund\inst{4},
S. Brandt\inst{4},
N.J. Westergaard\inst{4},
L. Amati\inst{1}, \\
E. Caroli\inst{1},
S. Del Sordo\inst{5},
G. Di Cocco\inst{1},
P. Durouchoux\inst{6},
R. Farinelli\inst{7},
F. Frontera\inst{1,7},
M. Orlandini\inst{1} \\ 
and A. Zdziarski\inst{8}
}

\institute{
Istituto di Astrofisica Spaziale e Fisica Cosmica --- Sezione di Bologna, 
CNR, via Gobetti 101, I-40129, Bologna, Italy
\and
NASA Goddard Space Flight Center, Code 661, Greenbelt, MD 20771, USA
\and 
Joint Center for Astrophysics, Department of Physics, University of 
Maryland, Baltimore County, MD 21250, USA
\and
Danish Space Research Institute, DK-2100 Copenhagen 0, Denmark
\and
Istituto di Astrofisica Spaziale e Fisica Cosmica --- Sezione di Palermo,
CNR, via La Malfa 153, I-90146 Palermo, Italy
\and
CEA Saclay, DSM, DAPNIA, Service d'Astrophysique, 91191 Gif sur Yvette 
Cedex, France
\and
Dipartimento di Fisica, Universit\`a di Ferrara, via Paradiso 12, I-44100
Ferrara, Italy
\and
N. Copernicus Astronomical Center, Bartycka 18, 00-716 Warsaw, Poland
}

\offprints{N. Masetti (\texttt{masetti@bo.iasf.cnr.it)}}
\date{Received February 5, 2004; accepted May 11, 2004}

\abstract{
Here we report results of an {\it INTEGRAL}-AO1 observation of the X-ray
burst and atoll source Ser X-1 performed in May 2003. The object was
observed for a total on-source time of 400 ks but nearly 8$^{\circ}$
off-axis due to its amalgamation with an observation of SS 433, the
pointing target source. Ser X-1 has been clearly detected up to 30 keV
with unprecedented positional accuracy for high-energy emission. The
20--30 keV light curve showed substantial variability during the
observation. Comparison with previous observations indicates that the
source was in its high (`banana') state and displayed a soft spectrum
during the {\it INTEGRAL} pointing. A (non simultaneous)
radio-to-$\gamma$--rays broad-band spectral energy distribution is also 
presented for the first time and discussed.

}

\titlerunning{Ser X-1 observed with {\it INTEGRAL}}
\authorrunning{N. Masetti et al.}

\maketitle

\section{Introduction}

\begin{figure*}[t!]
\psfig{file=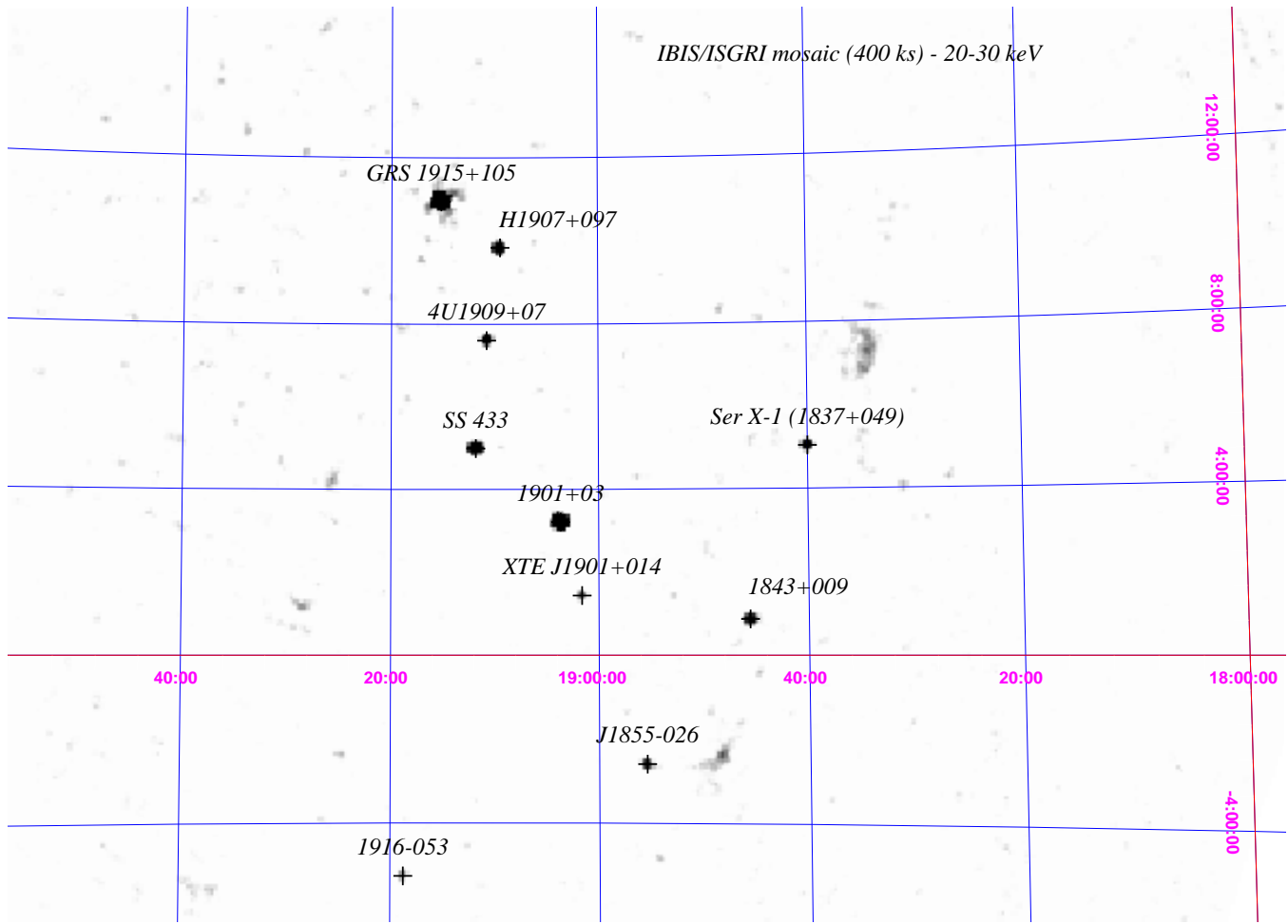,width=18cm}
\caption{Part of the mosaic image of the Ser X-1 field as imaged by ISGRI 
in the 20--30 keV band and accumulated over the whole observation (400 ks
on-source time). The field is about 30$^{\circ}$$\times$25$^{\circ}$. 
North is up, East is left. Ser X-1 is detected at $\sim$25$\sigma$ 
significance. The image greyscale is histogram-equalization coded and 
scaled to the intensity, with the white color corresponding to zero counts.
The large unidentified features northwest of Ser X-1 and west of 
J1855$-$026 are artifacts (ghost residuals) produced by the bright source 
GRS 1915+105.}
\end{figure*}

Low Mass X-ray Binaries (LMXRBs) form the most populated class of
Galactic X-ray binaries. They are X-ray binary systems in which the mass
donor star is less than about 1 $M_\odot$ and loses matter via 
Roche-lobe overflow to a very compact accretor. Matter forms an accretion 
disk as it falls towards the compact object. Many of these systems
contain a weakly magnetized neutron star (NS) and are persistently 
bright in X-rays.

The study of spectral and temporal properties of NS LMXRBs in the X-ray
energy band below 20 keV has shown the existence of two subclasses of 
sources. These were identified by their difference in the
hardness ratio from two adjacent X-ray bands as a function of the source
intensity in the total energy band (hardness-intensity diagram) and/or
their track on an X-ray `colour-colour' diagram (CD) constructed by using 
two hardness ratios using different energy bands. These two subclasses were
labeled as Z sources, if they track a Z-shape, or part of it, in the CD
and atoll sources if their position in the CD draws an atoll-shaped
figure. Indeed, Z sources are in general permanently bright, while atoll
sources show flux variations by factors ranging from about 5 to about 1000
or more and, in parallel, substantial spectral variations (e.g., Piraino
et al. 1999). Also, the timing behaviour is different between the two
subclasses; see e.g. van der Klis (2000) for details on this 
classification and on the properties of these subclasses. However, recent 
results indicate that atoll sources actually track a Z shape on year-long 
timescales (Gierli\'nski \& Done 2002; Muno et al. 2002).

Ser X-1, or 4U 1837+04, has been classified as an atoll source (e.g., Liu 
et al. 2001). Archival {\it EXOSAT} data (Seon \& Min 2002) showed this
object in the `banana' (i.e. high luminosity) state during the observations.
More recent {\it BeppoSAX} and {\it RXTE} pointings (Oosterbroek et al.
2001) caught the source while it was again in a high activity state, with 
an unabsorbed flux (1--200 keV) of 8.0$\times$10$^{-9}$ erg cm$^{-2}$ 
s$^{-1}$. We refer the reader to the work of these authors for the 
most recent description of the broadband X--ray spectral and timing 
properties of this source.
Ser X-1 has never been seen in the hard (`island') state that is 
observed in atoll sources at low luminosities (Hasinger \& van der Klis 
1989).

This source also exhibits Type I X-ray bursts (e.g. Lewin et al. 1995);
moreover, during 2001, {\it BeppoSAX} pinpointed a very long ($\sim$4
hours) X-ray burst (Cornelisse et al.  2002), making this source join the
group of `superbursters' (see Kuulkers 2003 for a review). By studying
X-ray bursts observed with {\it Einstein}, Christian \& Swank (1997)
deduced a distance to the source of 8.4 kpc. This implies a 1--200 keV
luminosity of 6.7$\times$10$^{37}$ erg s$^{-1}$ during the {\it BeppoSAX}
observation of Oosterbroek et al. (2001), which means roughly one third of
the Eddington luminosity for a NS.

The optical counterpart to Ser X-1, located in a crowded stellar field
(Thorstensen et al. 1980), was correctly identified by Wachter (1997) and,
subsequently, spectroscopically confirmed and studied by Hynes et al.
(2004). Very recently, the radio counterpart was discovered with the VLA
(Migliari et al. 2004).

In this paper we report and discuss the results of an observation of 
Ser X-1 performed with the INTErnational Gamma--RAy Laboratory ({\it 
INTEGRAL}; Winkler et al. 2003) during AO1 on May 2003, i.e. less than 7 
months after the launch of this spacecraft.
This paper is thus structured as follows: Sect. 2 describes the 
observations and the data analysis, Sect. 3 reports the results of the 
{\it INTEGRAL} pointing as well as the construction of a broadband 
spectral energy distribution (SED) for Ser X-1 using literature data 
along with the high-energy ones presented here; then, in Sect. 4 a 
discussion will be given.

\section{Observations and data analysis}

The observations presented here were acquired during revolutions 67 to 69,
i.e. between 12:00 UT of 3 May 2003 and 09:26 UT of 9 May 2003, for a
total on-source time of 400 ks, with the IBIS (Ubertini et al. 2003) 
and SPI (Vedrenne et al. 2003) instruments onboard {\it INTEGRAL}. These
detectors allow an actual 20 keV -- 10 MeV spectral coverage altogether. 
In particular, the IBIS telescope is composed of two detector layers: 
ISGRI (Lebrun et al. 2003), optimized for the energy range 20--200 keV, 
and PICsIT (Di Cocco et al. 2003; Labanti et al. 2003), covering the range
from 175 keV to 10 MeV. 
Data were acquired with a spacecraft rectangular 5$\times$5 dithering 
pattern mode. 

The field of Ser X-1 was also intermittently observed with Unit 2 of the 
coded-mask X-ray monitor JEM-X on board {\it INTEGRAL} (Lund et al. 
2003). 
During the period in which the observation 
reported here was performed, Unit 1 of JEM-X was not available,
being turned off to avoid fast degradation of the detector.
The present {\it INTEGRAL} observation of Ser X-1 was amalgamated 
with one targeting SS433 (Cherepashchuk et al. 2003), thus leaving Ser X-1
about 7$\fdg$8 off axis; hence, JEM-X could only observe this object
for a small fraction of the total 400 ks exposure time. All JEM-X 
observations of Ser X-1 were at an off-axis angle larger than 3$\fdg$6; 
we only considered JEM-X pointings in which the source was less 
than 4$\fdg$5 off axis, because of uncertainties in the knowledge of 
the instrument response at larger offset angles. Thus, JEM-X data from 
22 pointings, corresponding to about 40 ks of observation time, have 
been used in the present analysis.

Ser X-1 was outside the field of view of the fourth instrument carried by
{\it INTEGRAL}, i.e. the optical camera OMC (Mas-Hesse et al. 2003), and
therefore no data collected with OMC were available for this source during
the present {\it INTEGRAL} pointing.

The data reduction and analysis was performed with the standard Offline
Scientific Analysis (OSA v3.0\footnote{Available through the INTEGRAL
Science Data Centre (ISDC, Geneva, Switzerland) at the web address:
{\tt http://isdc.unige.ch/index.cgi?Soft+download}}), whose algorithms are 
described in Goldwurm et al. (2003a) for IBIS and in Diehl et al. (2003) 
for SPI. The count rates extracted from the standard pipeline were 
normalized to those measured during off-axis observations of the Crab, 
in order to obtain fluxes in terms of physical units.
Given the present uncertainties in the flux evaluation for off-axis 
sources (see Goldwurm et al. 2003b), we adopted a procedure similar to 
that explained by Goldoni et al. (2003) to analyse the IBIS data for 
off-axis sources. For JEM-X the analysis was performed using software 
still under development at DSRI, Copenhagen (Denmark). The count 
rates extracted for Ser X-1 were compared to results from off-axis Crab 
observations analyzed with the same software. For details concerning the 
JEM-X analysis software see Westergaard et al. (2003).

\section{Results}

\begin{table}
\caption[]{Fluxes and luminosities for Ser X-1 measured during the present 
{\it INTEGRAL} observation. For each measurement, the instrument and 
the energy band are reported. Quoted errors are at 1$\sigma$ confidence 
level, while upper limits are at 3$\sigma$ level. Luminosities are 
computed assuming a distance $d$ = 8.4 kpc.}
\begin{center}
\begin{tabular}{l|c|r|r}
\hline
\noalign{\smallskip}
\multicolumn{1}{c|}{Energy range} & {\it INTEGRAL} & 
\multicolumn{1}{c|}{Flux} & \multicolumn{1}{c}{Luminosity} \\
\multicolumn{1}{c|}{(keV)} & instrument & 
\multicolumn{1}{c|}{(erg cm$^{-2}$ s$^{-1}$)} & 
\multicolumn{1}{c}{(erg s$^{-1}$)} \\ 
\noalign{\smallskip}
\hline
\noalign{\smallskip}
3--5          & JEM-X          & (2.2$\pm$0.2)$\times$10$^{-9}$ &
	1.9$\times$10$^{37}$ \\
5--10         & JEM-X          & (1.76$\pm$0.15)$\times$10$^{-9}$ &
	1.5$\times$10$^{37}$ \\
10--15        & JEM-X          & (0.44$\pm$0.10)$\times$10$^{-9}$ &
	3.7$\times$10$^{36}$ \\
15--35        & JEM-X          & $<$1$\times$10$^{-9}$ &
	$<$8.4$\times$10$^{36}$ \\
20--30        & ISGRI          & (6.2$\pm$0.3)$\times$10$^{-11}$ &
	5.2$\times$10$^{35}$ \\
20--30        & SPI            & $<$5.0$\times$10$^{-11}$ &
	$<$4.2$\times$10$^{35}$ \\
30--40        & ISGRI          & $<$6$\times$10$^{-12}$ &
	$<$5$\times$10$^{34}$ \\
30--40        & SPI            & $<$2.6$\times$10$^{-11}$ &
	$<$2.2$\times$10$^{35}$ \\
220--280      & PICsIT         & $<$1$\times$10$^{-10}$ &
	$<$8$\times$10$^{35}$ \\
\noalign{\smallskip}
\hline
\end{tabular}
\end{center}
\end{table}

Measurements and upper limits for the X-ray emission in various 
energy bands from Ser X-1 during the present {\it INTEGRAL} pointing are 
listed in Table 1. Below we report the findings obtained with each 
instrument onboard {\it INTEGRAL}.

Ser X-1 was clearly detected with ISGRI in the 20--30 keV band (Fig. 1)
with a significance of $\sim$25$\sigma$ over the whole 400 ks observation,
with a count rate of 0.84$\pm$0.03 counts s$^{-1}$.
Nothing was detected above 30 keV at the position of this source: the 
upper limit to the emission in the 30--40 keV range is reported in Table 
1. Our ISGRI results agree with those obtained by Molkov et al. (2004) 
using {\it INTEGRAL} observations acquired before and after ours.

The object was also not detected in the whole spectral range covered by
PICsIT. A $3\sigma$ upper limit of 4$\times 10^{-6}$ photons cm$^{-2}$
s$^{-1}$ in the energy band centered at $250$ keV and $\Delta E$ = 60 keV
wide was obtained from the PICsIT data (see Table 1).  

SPI also did not reveal any significant emission from Ser X-1.
From Table 1 one can note an apparent discrepancy between the flux 
obtained from the ISGRI detection and the SPI upper limit: this can be 
explained in terms of a not yet complete correction for systematic errors.

JEM-X detected the source in the 3--15 keV band with significance 
$\sim$30$\sigma$ in 2200 s when the source was 3$\fdg$6 off axis.
Using off-axis observations of the Crab (as mentioned in the previous 
section), we measured, in the three bands 3--5, 5--10 and 10--15 
keV, the fluxes shown in Table 1. A 3$\sigma$ upper limit was instead 
obtained from the 15--35 keV band data; its value is in agreement 
with the ISGRI 20--30 keV detection.

The ISGRI source detection was at coordinates $\alpha$ =  18$^{\rm h}$ 
40$^{\rm m}$ 00$\fs$0, $\delta$ = +05$^{\circ}$ 02$'$ 06$''$ (J2000). 
The significance of the detection implies a 90\% confidence positional 
accuracy of 1$'$ (Gros et al. 2003).
As regards JEM-X, it detected Ser X-1 at $\alpha$ = 18$^{\rm h}$ 
39$^{\rm m}$ 56$\fs$2, $\delta$ = +05$^{\circ}$ 01$'$ 57$''$ (J2000)
with a 99\% confidence uncertainty of 40$''$ in both coordinates.
The two positions are in good agreement with each other, given the 
off-axis location of the source within the fields of view of ISGRI and 
JEM-X. The object localization afforded by these two instruments onboard 
{\it INTEGRAL} is moreover fully consistent with that of the optical 
counterpart MM Ser (Thorstensen et al. 1980; Wachter 1997) as well as with 
the radio position (Migliari et al. 2004).

The 20--30 keV light curve of Ser X-1, rebinned to $\sim$40 ks, is shown 
in Fig. 2. We chose this temporal resolution as it turned out to be the best 
tradeoff between the time sampling and the S/N ratio of each bin.
Substantial fluctuations are apparent in this band; so, in order 
to test their reality, we performed a fit to the data points on the 
hypothesis that the emission is constant. By means of a standard 
$\chi^{2}$ test it was found that the constancy of the 20--30 keV 
emission from this source is rejected at the 99.99\% confidence level.

Figure 2 also shows the Ser X-1 light curves in the 3--5, 5--10 and 10--15
keV ranges as obtained with JEM-X. Each point corresponds to a
single JEM-X pointing ($\sim$2 ks long).
We also explored the JEM-X light curves using higher time resolutions down
to 500 s in the parts of the observation covered by the instrument (slots
lasting $\sim$5 ks on average and separated by off-source intervals of
about 20 ks). We did not find any hint of X-ray bursts or superbursts in
these light curves. We of course cannot exclude the possibility that X-ray
(super)bursts from Ser X-1 may have occurred in the 6 days covered by our
{\it INTEGRAL} observation during the time intervals in which the source 
was outside the JEM-X field of view.

Observations with the All-Sky Monitor (ASM\footnote{ASM light curves are
available at:\\ {\tt http://xte.mit.edu/ASM\_lc.html}}; Levine et al.
1996) onboard the satellite {\it RXTE} (Bradt et al. 1993) showed that  
Ser X-1 had an average flux of 208$\pm$8 mCrab in the 1.5--12 keV band at 
the time of the {\it INTEGRAL} pointing, taking into account that 75 ASM 
counts s$^{-1}$ correspond to 1 Crab in this band. For comparison, during 
the {\it BeppoSAX} observation, the ASM data indicate that the source was 
at 239$\pm$3 mCrab in the same energy band.
This $\sim$15\% difference in brightness apparently was not reflected in 
the overall spectral shape in the 1.5--12 keV range: indeed, the ASM hardness 
ratios HR1 and HR2 (defined as the ratio between the ASM counts in the 
3--5 keV band and those in the 1.5--3 keV band, and as the ratio of the 
counts in the 5--12 keV band to those in the 3--5 keV band, respectively), 
were 1.01$\pm$0.11 and 1.20$\pm$0.10 during the {\it INTEGRAL} pointing 
and 1.02$\pm$0.04 and 1.20$\pm$0.03 at the time of the {\it BeppoSAX} 
observation, respectively.

In addition, the JEM-X fluxes measured from Ser X-1 during the present
{\it INTEGRAL} observation were found to be in general agreement with
those obtained with the ASM in the spectral range over which the
sensitivities of these two instruments overlap.

Feeling justified by these results, we expanded the spectral coverage of 
Ser X-1 down to 1.5 keV by adding to the {\it INTEGRAL} data set 
simultaneous 1.5--3 keV data acquired by the ASM.
These ASM counts in the 1.5--3 keV were thus averaged over the duration of 
the {\it INTEGRAL} pointing; next, they were converted into fluxes 
assuming the spectral description for Ser X-1 provided by Oosterbroek et 
al. (2001), care being taken to rescale it with the use of the archival 
ASM data obtained simultaneously with the {\it BeppoSAX} observation.

We then collected published optical data (Wachter 1997) and near-infrared 
(NIR) 2MASS data (Skrutskie et al. 1997), as well as far-infrared upper 
limits (Beichman et al. 1988) and radio data (Migliari et al. 2004; 
Wendker 1995 and references therein). This allowed us to construct, for 
the first time for Ser X-1, a broadband SED (Fig. 3) spanning radio to 
$\gamma$--ray frequencies. 
Although non-simultaneous with the high-energy part of the spectrum, these
lower frequency data and upper limits can give us a general description of
the overall SED of this source thanks to the relative long-term stability 
of its emission with time.

Optical and NIR data were corrected assuming a color excess $E(B-V)$ = 0.8
(e.g. Hynes et al. 2004 and references therein) and converted into fluxes
using the normalizations by Fukugita et al. (1995) for the optical and
the ones referring to the 2MASS\footnote{These normalizations are 
available at: \\
{\tt http://www.ipac.caltech.edu/2mass/releases/ \\ allsky/faq.html}}
for the NIR. Given the resolution of the 2MASS survey (pixel size: 2$''$) 
and the crowding of the Ser X-1 field (e.g. Wachter 1997), we considered 
the 2MASS NIR detections as conservative upper limits to the actual source 
fluxes in the $JHK$ bands.

\begin{figure}
\psfig{file=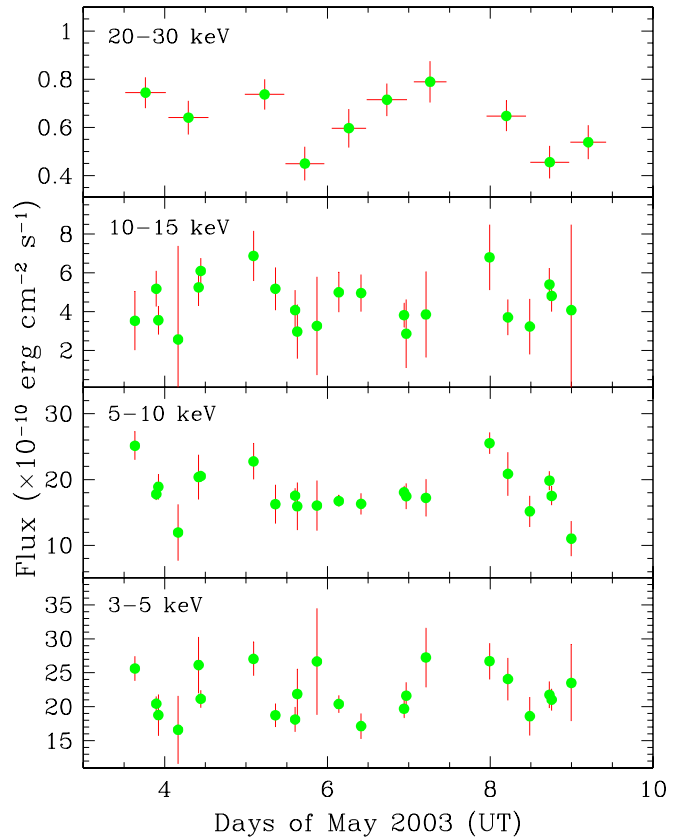,width=9cm}
\caption{Light curves of Ser X-1 as observed by ISGRI (20--30 keV) and 
JEM-X (3--5 keV, 5--10 keV and 10--15 keV) onboard {\it INTEGRAL} during 
the pointing presented in this paper. ISGRI data are rebinned to 
$\sim$40 ks, while JEM-X data are accumulated over each single 
pointing (with duration $\sim$2 ks).}
\end{figure}

\begin{figure} 
\hspace{-0.5cm}
\psfig{file=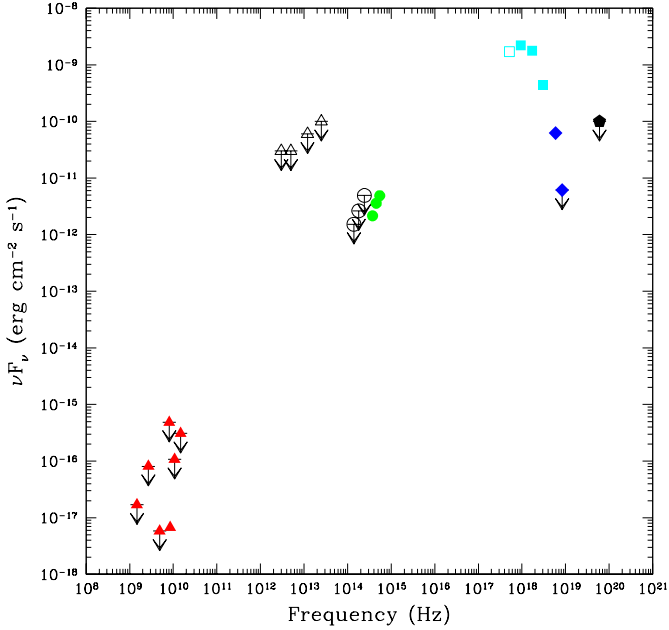,width=9.5cm} 
\caption{Broadband SED of Ser X-1 constructed with simultaneous data 
and upper limits from ISGRI (filled diamonds), PICsIT (filled pentagon), 
JEM-X (filled squares) and ASM data (open square) plus non-simultaneous 
optical measurements (filled dots), NIR and far-infrared upper limits 
(open circles and open triangles, respectively) and radio data (filled 
triangles). Error bars of data points are not reported as their size is 
comparable to, or smaller than, that of the symbols used in the figure. 
}
\end{figure}

\section{Discussion}

We observed Ser X-1 with {\it INTEGRAL}; we detected the source in the 
3--15 keV band with the JEM-X instrument, and in the 20--30 keV range
with the ISGRI detector of the IBIS instrument. Upper 
limits for the source flux were instead obtained in the 30--40 keV 
band and around 250 keV with the ISGRI and PICsIT detectors, 
respectively. The positional accuracy in the 20--30 keV band (1$'$) is 
unprecedented, and the observation indicates that no significant nearby 
hard source is present within a radius of few degrees from Ser X-1.
Thus we can exclude any contamination from persistent field sources in 
past observations of Ser X-1 made with previous non-imaging hard X-ray 
detectors (e.g., the HPGSPC and the PDS onboard {\it BeppoSAX}).

The source appears to be variable by a factor two in intensity in the
20--30 keV band on timescales of $\sim$100 ks (Fig. 2). A similar
behaviour was present in the 2--10 keV band as observed by the MECS
instrument onboard {\it BeppoSAX} (Oosterbroek et al. 2001), although in
this case the variability amplitude was about 25\%, thus not as strong as
in the {\it INTEGRAL} observations. This however is most likely a sampling
effect due to the shorter duration ($\sim$60 ks) of the {\it BeppoSAX}
pointing: indeed, during the first 80 ks of ISGRI data the flux variation
in the 20--30 keV band is comparable to that seen with {\it BeppoSAX}. 

The difference between the ISGRI and MECS light curves might however be
due to a further reason: as the 20--30 keV flux is likely dominated by the
Comptonization component, while that between 2 and 10 keV by the blackbody
(or disk-blackbody) originating close to the NS surface, the two
components might display variability with different timescales which in
turn may be reflected in the light curve shape of the two above mentioned
bands. Unfortunately, again the ASM and JEM-X light curves do not carry
enough information to confirm or disprove this hypothesis.

The light curve sampling of JEM-X and ISGRI did not allow us to look for 
X-ray bursts from the surface of the NS harboured in this system 
throughout the entire observation: only in the $\sim$40 ks covered by 
JEM-X can we exclude their presence. We therefore cannot rule out that 
some of the variability we see in the 20--30 keV range is caused by these 
phenomena occurring when JEM-X was pointing off-source; however, for 
typical values of their spectral shape, rate and overall energy output 
(see e.g. Sztajno et al. 1983), we do not expect that X-ray bursts can 
contribute substantially to the total energy radiated in 40 ks by Ser X-1 
in the 20--30 keV band, which amounts to $\sim$2$\times$10$^{40}$ erg. 

The observed $\sim$100 ks timescale variability also cannot be
explained assuming the occurrence of a `superburst' like the one reported 
by Cornelisse et al. (2002), which lasted $\sim$4 hours and emitted
$\approx$10$^{40}$ erg in the 20--30 keV range. Indeed such a phenomenon, 
if present, can be contained in a single 40-ks bin of the 20--30 keV light 
curve reported in Fig. 2. The variations detected between adjacent bins in 
this spectral range might instead have been produced by a superburst 
but, again, the sparse ASM dwell-by-dwell and JEM-X data coverage during 
the {\it INTEGRAL} pointing does not allow us to check this hypothesis.

It should moreover be noted that the trend seen in the 20--30 keV light
curve suggests that the source seems to vary periodically on a $\sim$3
days timescale: this roughly matches a possible (superorbital?)  X-ray
periodicity of 3.4 days already reported for Ser X-1 (Ponman 1981;  
Ritter \& Kolb 2003).

The 20--30 keV flux measurement obtained with {\it INTEGRAL} is about 20\%
lower than that obtained from the model by Oosterbroek et al. (2001)
in the same band (8$\times$10$^{-11}$ erg cm$^{-2}$ s$^{-1}$); this 
roughly agrees with the difference ($\sim$15\%) in the ASM flux as seen 
between the epochs of the {\it BeppoSAX} and {\it INTEGRAL} pointings.
The 30--40 keV to 20--30 keV flux ratio in the {\it INTEGRAL} data is 
$<$0.1 which is consistent with the value of 0.054 as measured by 
{\it BeppoSAX}.
The lack of sufficient coverage in the soft X-rays did not allow us to 
give a better description of the spectrum of this object; however, from 
the above results we can confidently say that we observed the atoll source 
Ser X-1 in its bright (banana) state, albeit at a lower intensity with 
respect to the time in which the {\it BeppoSAX} pointing was done, i.e., 
September 1999.

The broadband SED of Ser X-1 clearly shows a cut-off in the energy
distribution above 3$\times$10$^{18}$ Hz ($\simeq$10 keV). This can be
reasonably taken to be due to the drop of the Comptonization tail induced 
by the electron temperature, $kT_{\rm e^{-}} \sim$ 3 keV as determined by
Oosterbroek et al. (2001), and to the lack of further components in the
$\gamma$--ray domain. No simple model can fit the overall SED: while the 
optical data fall above the extrapolation of the {\it BeppoSAX} X-ray 
model by a factor $\sim$100, a smooth connection between optical and radio 
data may be present. 
Indeed, radio and optical fluxes are marginally consistent with lying on 
the same power law, $F_\nu \propto \nu\,^{\beta}$, with $\beta \sim$ 0.2.
This might indicate that the mechanism responsible for the radio
emission (possibly a jet) can play a substantial role in the observed 
optical light from this source also (see e.g. Fender 2003 and Migliari 
et al. 2004).

\begin{acknowledgements}
We thank Simone Migliari for having communicated to us the Ser X-1 radio 
detection result prior to publication. Pavel Binko is acknowledged for 
the help in the {\it INTEGRAL} data retrieval from the ISDC archive.
This work has made use of the NASA Astrophysics Data System Abstract 
Service, of the SIMBAD database, operated at CDS, Strasbourg, France, 
and of data products from the 2MASS.
ASM data were provided by the {\it RXTE} ASM teams at MIT and at the 
{\it RXTE} SOF and GOF at NASA's GSFC. This research was partially funded 
by ASI. We also thank the referee for several comments which helped 
us to improve this paper.
\end{acknowledgements}


\end{document}